\documentstyle[12pt,aasms4]{article}

\def\lsim{\lower.5ex\hbox{$\; \buildrel < \over \sim \;$}}
\def\gsim{\lower.5ex\hbox{$\; \buildrel > \over \sim \;$}}

\makeatletter
\def\jnl@aj{AJ}
\ifx\revtex@jnl\jnl@aj\let\tablebreak=&&&\nl\fi
\makeatother

\slugcomment{Revised version submitted to the ApJ Letters 20 Apr 1999}
 
\begin{document}

\title{Is SGR 1900+14 a Magnetar?}
 
\author{D. Marsden, R. E. Rothschild, and  R. E. Lingenfelter}
\affil{Center for Astrophysics and Space Sciences, University of 
California at San Diego \\ La Jolla, CA 92093} 

\begin{abstract}

We present {\it RXTE} observations of the soft gamma--ray repeater 
SGR 1900+14 taken September 4-18, 1996, nearly 2 years before the 
1998 active period of the source. The pulsar period (P) of $5.1558199
\pm 0.0000029$ s and period derivative ($\dot{P}$) of $(6.0\pm 1.0)
\times 10^{-11}$ s s$^{-1}$ measured during the 2-week observation 
are consistent with the mean $\dot{P}$ of $(6.126\pm 0.006)\times 
10^{-11}$ s s$^{-1}$ over the time up to the commencement of the 
active period. This $\dot{P}$ is less than half that of $(12.77\pm 
0.01)\times 10^{-11}$ s s$^{-1}$ observed during and after the active 
period. If magnetic dipole radiation were the primary cause of the 
pulsar spindown, the implied neutron star magnetic field would exceed 
the critical field of $\approx 4.4\times 10^{13}$ G by more than an 
order of magnitude, and such field estimates for this and other SGRs 
have been offered as evidence that the SGRs are magnetars, in which 
the neutron star magnetic energy exceeds the rotational energy. The 
observed doubling of $\dot{P}$, however, would suggest that the pulsar 
magnetic field energy increased by more than $100\%$ as the source 
entered an active phase, which seems very hard to reconcile with 
models in which the SGR bursts are powered by the release of magnetic 
energy. Because of this, we suggest that the spindown of SGR 1900+14 
is not driven by magnetic dipole radiation, but by some other process, 
most likely a relativistic wind. The $\dot{P}$, therefore, does not 
provide a measure of the pulsar magnetic field strength, nor evidence 
for a magnetar.
\end{abstract}
 

\section{Introduction}
\label{intro}

Soft gamma--ray repeaters (SGRs) are a class of astrophysical 
sources that emit bursts of high energy x--ray and gamma--ray 
radiation which are among the most energetic events in the Galaxy. 
The apparent association of their positions with supernova remnants 
and the detection of pulse periods in their nonbursting emission 
strongly suggest that the SGRs are young neutron stars (e.g. Mazets 
et al. 1979, and review by Rothschild 1995). The SGRs may also be 
related to the anomalous X--ray pulsars (AXPs: \cite{mer97}), which 
have comparable long ($>$ few second) periods. The observed SGR burst 
energies, assuming isotropic emission, range from typical values of 
$\sim 10^{41}$ ergs to as much as $10^{44}$ ergs in rare giant flares, 
such as that of 5 March 1979 from the SGR 0529--66 in the Large 
Magellanic Cloud. Suggested energy sources for these bursts have 
included, i) the rotational energy of the neutron star, $\sim 10^{45}
{(P/3.1\,\rm{s})}^{-2}$ ergs, where $P$ is the spin period, which might 
be tapped by pulsar glitches (e.g. \cite{baym71}), ii) the magnetic 
field energy $\sim 10^{44}{(B/B_{q})}^{2}$ ergs of {\it magnetars} with 
surface magnetic fields much greater than the quantum critical field 
$B_{q}=m_{e}^{2}c^{3}/ e\hbar\approx 4.4\times 10^{13}$ G tapped by 
magnetic-stress driven crustal quakes and magnetic reconnection 
(\cite{thompson95}), and iii) the gravitational binding energy of 
the neutron star, $\sim 10^{53}$ ergs, tapped by quakes (e.g Ramaty 
et al. 1980), and driven by plate tectonics (\cite{ruderman91}). 

Recent measurements of the rapid spindown rates of the SGR pulsars 
have been taken (e.g. \cite{kouv98}, 1999) as evidence for the 
magnetar hypothesis, in which the magnetic energy of the neutron 
star exceeds the rotational energy. Pulsations have been observed 
from three of the SGRs: SGR 0526--66 ($8$ s: \cite{mazets79}), 
SGR 1806--20 ($7.47$ s: \cite{kouv98}), and SGR 1900+14 ($5.16$ 
s: \cite{hurley99b}). The period derivatives ($\dot{P}$) of these 
pulsars have been found by either direct measurement (SGRs 1806--20 
and 1900+14) or by $\dot{P}=0.5 P/t_{snr}$, where $P$ is the pulse 
period and $t_{snr}$ is the estimated age of the associated supernova 
remnant (SGR 0526--66). If the spindown is driven by magnetic dipole 
radiation from an orthogonally rotating vacuum magnetic dipole, it 
can be shown (\cite{pacini68}) that the surface magnetic field is 
given by $B_{0}\approx 3.2\times 10^{19}\sqrt{P\dot{P}}$ G, which 
would yield surface magnetic fields of $6\times 10^{14}$, $8\times 
10^{14}$, and $5\times 10^{14}$ G for SGRs 0526--66 
(\cite{thompson95}), 1806--20 (\cite{kouv98}), and 1900+14 
(\cite{kouv99}), respectively. Here we present {\it RXTE} 
observations, however, which suggest that the spindown rate 
of SGR 1900+14 is due to torques other than those provided by 
the magnetic field, and thus does not provide evidence of a 
supercritical surface dipole field.

\section{Observations \& Analysis}
\label{obervations}

SGR 1900+14 was observed by the Proportional Counter Array (PCA) 
and High Energy X--ray Timing Experiment (HEXTE) instruments aboard 
the {\it Rossi X--ray Timing Explorer} on a number of occasions 
during the period September 4-18, 1996. The total exposure time was 
$\sim 47$ ks, with a temporal baseline of $15.4$ days. For the 
first $22$ ks, {\it RXTE} was pointed at a position RA (J2000)$= 
286^{\circ}.82$ and Dec (J2000)$=9^{\circ}.32$, which is $\sim 48 
\arcsec$ from the precise VLA position of SGR 1900+14 (\cite{frail99}), 
but well inside the $1^{\circ}$ FWHM field of view of the {\it RXTE} 
pointed instruments. Midway through the observations, the pointing 
position was changed to exclude the bright $438$ s binary x--ray 
pulsar 4U 1907+09 (\cite{zand98}) from the field of view. The 
second half of the observation ($25$ ks) was then conducted at 
the pointing position RA$=286^{\circ}.43$ and Dec$=8^{\circ}.98$, 
which is $\sim 0^{\circ}.35$ from the position of the SGR. As luck 
would have it, this field also contained a relatively bright 
confusing source, the $89$ s transient x--ray pulsar XTE J1906+09, 
which was discovered during the observation ({\cite{marsden98}). 
Finally, the Galactic Ridge emission is also a significant contributor 
to the x--ray flux in the {\it RXTE} field of view (\cite{valinia98}), 
due to the low Galactic latitude of SGR 1900+14 ($b\sim 0^{\circ}.75$). 
Because of these complications, we do not attempt to determine the 
x--ray spectrum of the SGR with the {\it RXTE} data, and instead 
concentrate on the temporal analysis. For information on the x--ray 
spectrum of the source, the reader is referred to Hurley et al. 
(1999b), Kouveliotou et al. (1999), and Murakami et al. (1999).  

The pointed x--ray instruments aboard {\it RXTE} are the High Energy 
X--ray Timing Experiment (HEXTE) and the Proportional Counter Array 
(PCA). HEXTE consists of two clusters of collimated NaI/CsI phoswich 
detectors with a total net area of $\sim 1600$ cm$^{2}$ and and 
effective energy range of $\sim 15-250$ keV (\cite{rothschild98}). 
The PCA instrument consists of five collimated Xenon proportional 
counter detectors with a total net area of $7000$ cm$^{2}$ and an 
effective energy range of $2-60$ keV (\cite{jahoda96}). The uncertainty 
in the timing of x--ray photons by the PCA and HEXTE is $<<1$ ms (\cite{rots98}), and is therefore negligible in the temporal analysis 
presented here.

The PCA and HEXTE photon times were corrected to the Solar System 
barycenter using the JPL DE200 ephemeris and the SGR coordinates RA(J2000)$=19^{\rm{h}} 07^{\rm{m}} 14\fs 33$ and Dec(J2000)$=+09^{\circ}19\arcmin 20\arcsec.1$ (\cite{frail99}). 
The PCA data were searched for pulsations using the chi-squared 
folding method, which calculates the value of chi-squared for a 
pulsar lightcurve (versus a constant rate) folded on a range of 
trial pulsar periods. Here the pulse phase $\phi$ for a given 
photon time $t$ is defined by the relation $\phi(t)=f(t-t_{0})+
{1\over 2}\dot{f}{(t-t_{0})}^{2}$, where the pulsar frequency $f$ 
and frequency derivative $\dot{f}$ are related to the period $P$ 
and period derivative $\dot{P}$ by the expressions $P=1/f$ and 
$\dot{P}=-\dot{f}P^{2}$. A maximum value of chi-squared occurs 
when the data are folded on the true pulsar period and period 
derivative. 

The PCA data were initially searched for pulsations using a range 
of $\sim 500$ periods about $5.153642$ s, the SGR 1900+14 period 
predicted from the timing ephemeris given in Kouveliotou et al. 
(1999). A significant chi-squared peak was seen, and a finer search 
was then conducted on a grid in $P-\dot{P}$ space around the peak, for 
a broad range of $\dot{P}$ including the value of $\dot{P}\sim 10^{-10}$ 
s s$^{-1}$ found by Kouveliotou et al. (1999). The results of the grid 
search are shown in Figure $1$. To estimate the confidence regions of 
$P$ and $\dot{P}$ indicated by the peak in chi-squared, we folded the 
$2-10$ keV PCA data with $P$ ($\dot{P}$) values slightly displaced from 
the peak value, while holding $\dot{P}$ ($P$) fixed at its peak value. 
The resultant lightcurves were then compared to a template lightcurve 
using the chi-squared test, and the $90\%$ confidence contours were 
calculated using the chi-squared probability distribution. A folding 
time midway through the {\it RXTE} observation was used throughout 
the analysis to minimize correlations between $P$ and $\dot{P}$.   
 
Using this analysis, we obtain a timing solution of $P=5.1558199\pm 
0.0000029$ s and $\dot{P}=(6.0\pm 1.0)\times 10^{-11}$ s s$^{-1}$, 
referenced to $t_{0}=50338.216$ (MJD). The errors are $90\%$ confidence. 
A search of the $15-100$ keV HEXTE data for the pulsar, using the PCA 
timing solution, failed to produce evidence of significant pulsations, 
which is not surprising given the faintness of the source and the 
presence of the bright confusing sources. The folded SGR 1900+14 
pulsar lightcurve for three PCA energy ranges, using the above 
timing parameters, is shown in Figure $2$. The pulsed fraction of 
the SGR 1900+14 is not constrained by these data, due to the uncertain 
x--ray flux from XTE J1906+09, 4U 1907+09, and the Galactic Ridge in 
the {\it RXTE} bandpass.

\section{Discussion}
\label{discussion}

The $2-10$ keV SGR 1900+14 lightcurve obtained here is virtually 
identical to the lightcurves obtained just before (\cite{hurley99b}) 
and just after (\cite{kouv99}) the commencement of the May 1999 active 
period of the source. This indicates that the x--ray emitting geometry 
is stable on timescales of years while the source is inactive. The 
lightcurve appears to have multiple components which vary differently 
with energy. There are three peaks in the $2-10$ keV lightcurve, with 
a single relatively broad central peak surrounded by two narrower peaks. 
The narrow peaks have harder spectra than the broad peak, as the narrow 
peak emission dominates the emission from the broad peak above $10$ keV. 
A simple explanation for the lightcurve morphology is that the pulsed 
emission consists of different emission components arising from 
different regions of the stellar surface. The narrow components may 
be beamed emission from a collimated wind off of relatively small 
hotspots, while the broader component could be more isotropic emission 
from a larger and cooler area of the crust. The two narrow components 
are greatly reduced in the pulsar lightcurves obtained just after the 
giant flare of August 27, 1999 (\cite{kouv99}; \cite{murakami99}), 
suggesting that the energy of the small hotspots may have been depleted 
during the active period.   

The observed temporal history of the SGR 1900+14 pulsar is shown in 
Figure $3$. The additional timing parameters of the present observations 
are important because they constrain the pulsar parameters long before 
the source went into outburst. Although the temporal coverage is 
incomplete, the {\it secular} spindown rate seems to change abruptly 
sometime close to the initiation of bursting, at which point the 
spindown continues steadily at an increased rate. These two different 
spindown rates are denoted by the dotted lines in Figure $3$, which are 
linear fits to the data before the outburst [up to and including the 
first observation of Kouveliotou et al. (1999)] and the data during 
and after the outburst [beginning with the first observation of 
Kouveliotou et al. (1999) and ending with the Shitov (1999) 
observation]. The third data point in Figure $3$, from Kouveliotou 
et al. (1999), appears to be near the change point in the spindown 
behavior because the period is consistent with the extrapolation of 
the pre-outburst timing solution, yet the $\dot{P}$ value measured 
during this observation is consistent with the outburst values. The 
fit to the data taken during and after the outburst period yields a 
value of $\dot{P}=(12.77\pm 0.01)\times 10^{-11}$ s s$^{-1}$ for the 
mean spindown rate, and the corresponding pre-outburst value is 
$\dot{P}=(6.126\pm 0.006)\times 10^{-11}$ s s$^{-1}$. Using these 
mean $\dot{P}$ values, the mean inferred dipole field strengths before 
and after the initiation of bursting would be $5.7\times 10^{14}$ G 
and $8.2\times 10^{14}$ G, respectively, if the spindown were driven 
by dipole radiation losses. These two values, which differ to a high 
degree of significance, would imply an abrupt increase in the SGR 
1900+14 magnetic field energy of more than $100\%$ around the time 
the source started bursting, which is contrary to the predictions of 
models in which the bursting is dissipating magnetic field energy. 

This discrepancy clearly suggests that the SGR 1900+14 spindown is 
not dominated by magnetic dipole radiation, and that the observed value 
of $P\dot{P}$ provides no direct measurement of $B$, and no direct 
evidence for a magnetar. Instead, the measured values of $P$ and 
$\dot{P}$ suggest that the SGR spindown may be due to {\it winds}, 
if we take the pulsar age to be that of the associated (\cite{hurley99a}) supernova remnant G42.8+0.6. Assuming that the initial period of the 
pulsar was much smaller than it is now, and that the braking index is 
constant in time, the pulsar age $t_{age}=P/[(n-1)\dot{P}]$, where the 
braking index $n$ is $3$ for pure dipole radiation but much less ($n\sim 
1$) for spindown due to wind torques. Taking the estimated age of 
G42.8+0.6 to be $\sim 10^{4}$ yr (\cite{vasisht94}, \cite{hurley96}), 
we find that the braking index for SGR 1900+14 must be $\sim 1$, i.e. 
$n = 1 + 0.16/(t_{age}/10^4\,\rm{yr})$, which indicates that the pulsar 
spindown is dominated by winds. The remnant age would have to be an 
order of magnitude smaller in order for the braking index to be consistent 
with that of dipole radiation, and in addition such an age would require 
an unreasonably large pulsar velocity of $\sim 2.5\times10^4$ km s$^{-1}$ 
for it to have traversed from the center of the remnant to its present 
position, assuming a distance of 5 kpc (\cite{vasisht94}, \cite{hurley96}). 
Thus the observations provide strong evidence that torques due to wind 
emission, and not magnetic dipole torques, dominate the spindown dynamics 
of SGR 1900+14.     

The spindown behavior of SGR 1900+14 can be explained simply if we 
assume that the spindown is due almost entirely to wind emission, as 
was also considered by Kouveliotou et al. (1999). Possible mechanisms 
for the generation of this wind include thermal radiation from hotspots 
and Alfv\'{e}n wave emission (\cite{thompson98}). In this interpretation, 
the SGR emits a robust wind of particles and fields, both during bursting 
and quiescent intervals, which carries away angular momentum from the 
star. The emission of a relativistic wind produces an exponential 
spindown of the pulsar $\Omega(t)=\Omega_{0}\exp(-kt)$, where $k$ is 
a constant parameterizing the rotational energy loss rate due to the 
wind (\cite{thompson98}). Using this relation, and the values of $P$ 
and $\dot{P}$ from our observations, we obtain $k=\dot{P}/P\sim 
2700^{-1}$ yr$^{-1}$. Given an age of $(1-2)\times 10^{4}$ yr for 
G42.8+0.6, we obtain an initial pulsar spin period of $P_{0}\sim 
3-120$ ms for SGR 1900+14, which is similar to the spin periods of 
young isolated pulsars such as the Crab. This $P_{0}$ is most likely 
an upper limit, given the likelihood of active periods (with higher 
spindown rates) in the past. 

As mentioned above, one scenario is that the spindown of SGR 1900+14 
is due to Alfv\'{e}n wave emission, in which a stream of particles and 
fields escape the star along magnetic field lines forced open by the 
wind pressure (\cite{thompson98}). A supercritical magnetic field is 
not required for this mechanism to explain the SGR 1900+14 spindown. 
From Thompson \& Blaes (1998), the spindown constant is given by
\begin{equation}
k=1.5\times 10^{-11}{\left({B_{\ast}\over 3\times 10^{12}\,\rm{G}
}\right )}^{2}{\left({\delta B_{\ast}/B_{\ast}\over0.01}\right )}^{4/3}
\,\rm{Hz},
\end{equation}
where $B_{\ast}$ is the dipole field strength, $\delta B_{\ast}$ 
is the wave amplitude, and we have assumed a neutron star moment of 
inertia and radius of $1.1\times 10^{45}$ g cm$^{2}$ and $10$ km, 
respectively. This value of $k$ is comparable to the measured value 
$k=\dot{P}/P\sim 10^{-11}$ Hz for SGR 1900+14, indicating that this 
mechanism can explain the spindown of the SGR with conventional 
($\sim 10^{12}$ G) field strengths, assuming that there is a 
mechanism to continuously generate Alfv\'{e}n waves. 

Even though a supercritical magnetic field on a global scale can not 
account for the SGR pulsar spindown, such fields on much smaller 
localized scales may nevertheless play an important role in the 
bursting process. Since the wind torques initially operate to spin 
down the neutron star crust, one might expect that if the core is not 
rigidly coupled to the crust, then the core could be spinning slightly 
faster and the resulting differential rotation could wind up any magnetic 
field threading between the core and crust, building up large internal 
magnetic field pressures. By analogy to the Sun, we might expect that 
the growing pressure of the internal field is episodically released by 
the surface break out of intense magnetic fields in localized regions, 
similar to the appearance of sunspots, which have local fields of 10$^2$ 
to 10$^3$ times the average global surface field of the Sun. Such spots 
of emerging magnetic flux (EMF) on a neutron star may thus contain 
supercritical, or larger, localized fields, $B_s$ within radii $r_s$, 
with total magnetic energies $> 3\times 10^{41}(B_s/B_{q})^2(r_s/1\,
\rm{km})^3$ erg, and they may be accompanied by comparable tectonic 
stresses and heating from field diffusion in the crust. To contain 
the giant flare of August 27, 1999, for example, a local field with 
$B\sim B_{q}$ can contain the $3\times 10^{42}$ ergs of energy released (\cite{frail99}) within a bubble of radius $r_{s}\sim 2$ km, which is 
a small fraction of the surface area of the star. The occurrence of such 
EMF-spots could thus provide an episodic source of both magnetic and 
tectonic-gravitational energy release, both thermal and nonthermal, 
that power both the steady localized winds and the impulsive bursts 
of SGRs, much as the sunspot fields are dissipated in winds, flares 
and diffusion on the Sun. The solar analogy was also discussed by 
Sturrock (1986) for Galactic gamma--ray bursts.

The SGR wind hypothesis can also explain other observed features of 
the burst and quiescent emission from SGRs. If both the quiescent 
x--ray emission and the spindown torque of SGR 1900+14 are due to 
wind emission, the persistent x--ray flux and the spindown luminosity 
should be correlated (this is not true of SGR 1806--20, because of 
the surrounding plerion --- see below). Between the {\it ASCA} 
observations of Hurley et al. (1999b) and Murakami et al. (1999), 
the persistent x--ray flux of SGR 1900+14 increased by $(140\pm 20)
\%$. Using the appropriate mean $\dot{P}$ values from Figure $3$, 
the spindown luminosity increased by $\sim 120\%$ over the same time 
interval, which is consistent with the steady x--ray flux and spindown 
arising from the wind. 

The radio signature of SGR winds have been observed from SGRs 1900+14 
(\cite{frail99}) and 1806--20 (\cite{kulkarni94}). In the latter case, 
the SGR winds power a plerionic nebula with a total energy content 
($\sim 10^{45}$ ergs) much greater than the energy given off in 
a typical burst interval ($\sim 10^{43}$ ergs, \cite{kouv99}), 
explaining the lack of variability seen from the SGR 1806--20 x--ray 
and radio counterparts (\cite{sonobe94}; \cite{vasisht95}). In the 
case of SGR 1900+14, a {\it transient} wind nebula from relativistic 
particles injected during the giant flare of August 27, 1999 
(\cite{hurley99c}) was observed by the VLA (\cite{frail99}). The 
different radio properties of the SGR 1806--20 and SGR 1900+14 
counterparts are probably due to the different external pressures 
for the two sources, since SGR 1806--20 is still inside its high 
pressure SNR while SGR 1900+14 is outside its associated supernova 
remnant, where the confining pressure is relatively low. The weak 
confining pressure of SGR 1900+14 inhibits the formation of a bright 
plerion (\cite{frail99}). The observed nonthermal (photon index $\sim 
2.2$:  \cite{sonobe94}; \cite{hurley99b}) quiescent x--ray spectra of 
the active SGR sources is characteristic of emission from a magnetized 
wind (\cite{tavani94}). Finally, the burst spectra of SGRs can be 
explained by the Compton upscattering of soft photons in a mildly 
relativistic wind, without involving a supercritical stellar field (\cite{fat96}). 

\acknowledgments

We thank Duane Gruber for suggesting improvements in the timing analysis. 
This work was funded by NASA grant NAS5-30720.

\clearpage

\begin{figure}
\caption{~Determination of the SGR 1900+14 timing ephemeris. The grid 
of chi-squared values as a function of period and period derivative is 
shown for the $2-10$ keV PCA data. Shown are four linearly-spaced 
contours displaced from the peak by units of $\Delta {\chi}^{2}=20$. 
The dotted lines denote the $90\%$ confidence regions of $P$ and 
$\dot{P}$.} 
\end{figure}

\begin{figure}
\caption{~The SGR 1900+14 folded lightcurve. The pulsar lightcurve is 
shown for three different PCA energy bands.} 
\end{figure}

\begin{figure}
\caption{~The timing history of SGR 1900+14. The vertical dashed line 
indicates the approximate time at which the source entered a bursting 
phase, and the dotted lines indicate linear fits to the data up to 
the onset of bursting, and to the data after the onset.} 
\end{figure}


\begin{thebibliography}{}

\bibitem[Baym \& Pines 1971]{baym71} Baym, G., \& Pines, D. 1971, 
Ann. Phys., 66, 816

\bibitem[Fatuzzo \& Melia 1996]{fat96} Fatuzzo, M. \& Melia, F. 
1996, \apj, 464, 316

\bibitem[Frail, Kulkarni, \& Bloom 1999]{frail99} Frail, D. A., 
Kulkarni, S. R., \& Bloom, J. S. 1999, Nature, in press

\bibitem[Hurley et al. 1996]{hurley96} Hurley, K. et al. 1996, 
\apjl, 463, L13

\bibitem[Hurley et al. 1999a]{hurley99a} Hurley, K. et al. 1999a, 
\apjl, 510, L107

\bibitem[Hurley et al. 1999b]{hurley99b} Hurley, K. et al. 1999b, 
\apjl, 510, L111

\bibitem[Hurley et al. 1999c]{hurley99c} Hurley, K. et al. 1999c, 
Nature, 397, 41

\bibitem[in't Zand, Baykal, \& Strohmayer 1998]{zand98} in't Zand, 
J. J. M., Baykal, A., \& Strohmayer, T. E. 1998, \apj, 496, 386

\bibitem[Jahoda et al. 1996]{jahoda96} Jahoda, K. et al. 1996, 
EUV, X--ray, and Gamma--Ray Instrumentation for Astronomy VII, 
SPIE Proc. 2808, eds. O. H. V. Sigmund \& M. Gumm (Bellingham: 
SPIE), 59

\bibitem[Kouveliotou et al. 1998]{kouv98} Kouveliotou et al. 1998, 
Nature, 393, 235

\bibitem[Kouveliotou et al. 1999]{kouv99} Kouveliotou, C. et al. 1999, 
\apjl, 510, L115

\bibitem[Kulkarni et al. 1994]{kulkarni94} Kulkarni, S. R. et al. 
1994, Nature, 368, 129 

\bibitem[Marsden et al. 1998]{marsden98} Marsden, D. et al. 1998, 
\apjl, 502, L129

\bibitem[Mazets et al. 1979]{mazets79} Mazets, E. P. et al. 1979, 
Nature, 282, 587

\bibitem[Mereghetti, Stella, \& Israel 1998]{mer97} Mereghetti, 
S., Stella, L., \& Israel, G. L. 1998, in {\it The Active X--ray 
Sky: Results from BeppoSAX and Rossi-XTE}, Nuclear Physics B 
Proceeding Supplements, eds. L. Scarsi, H. Bradt, P. Giommi, \& 
F. Fiore, (Elsevier Science: New York), 253  

\bibitem[Murakami et al. 1999]{murakami99} Murakami, T. et al. 1999, 
\apjl, 510, L119

\bibitem[Pacini 1968]{pacini68} Pacini, F. 1968, Nature, 221, 

\bibitem[Ramaty et al. 1980]{ramaty80} Ramaty, R. et al. 1980, 
Nature, 287, 122

\bibitem[Rothschild 1995]{rothschild95} Rothschild, R. E. 1995 in 
{\it High Velocity Neutron Stars and Gamma--Ray Bursts}, AIP Conference 
Proceedings 384, eds R. E. Rothschild \& R. E. Lingenfelter (AIP 
Press: New York), 51

\bibitem[Rothschild et al. 1998]{rothschild98} Rothschild, R. E. et 
al. 1998, \apj, 496, 538

\bibitem[Rots et al. 1998]{rots98} Rots, A. H. et al. 1998, \apj, 
501, 749

\bibitem[Ruderman 1991]{ruderman91} Ruderman, M. 1991, \apj, 382, 
587

\bibitem[Shitov 1999]{shitov99} Shitov, Yu. P. 1999, IAUC 7001

\bibitem[Sonobe et al. 1994]{sonobe94} Sonobe, T., et al. 1994, \apjl, 
436, L23 

\bibitem[Sturrock 1986]{sturrock86} Sturrock, P. A. 1986, Nature, 321, 47

\bibitem[Tavani 1994]{tavani94} Tavani, M. 1994, \apj, 431, L83

\bibitem[Thompson \& Duncan 1995]{thompson95} Thompson, C. \& Duncan, 
R. C. 1995, \mnras, 275, 255

\bibitem[Thompson \& Blaes 1998]{thompson98} Thompson, C., \& Blaes, 
O. 1998, \prd, 57, 3219
 
\bibitem[Valinia \& Marshall 1998]{valinia98} Valinia, A. \& Marshall, 
F. E. 1998, \apj, 505, 134

\bibitem[Vasisht, Frail, Kulkarni, \& Greiner, 1994]{vasisht94} 
Vasisht, G., Frail, D. A., Kulkarni, S. R. \& Greiner, J. 1994, 
\apjl, 431, L35

\bibitem[Vasisht, Frail, \& Kulkarni 1995]{vasisht95} Vasisht, G., 
Frail, D. A., \& Kulkarni, S. R. 1995, \apjl, 440, L65

\end{thebibliography}
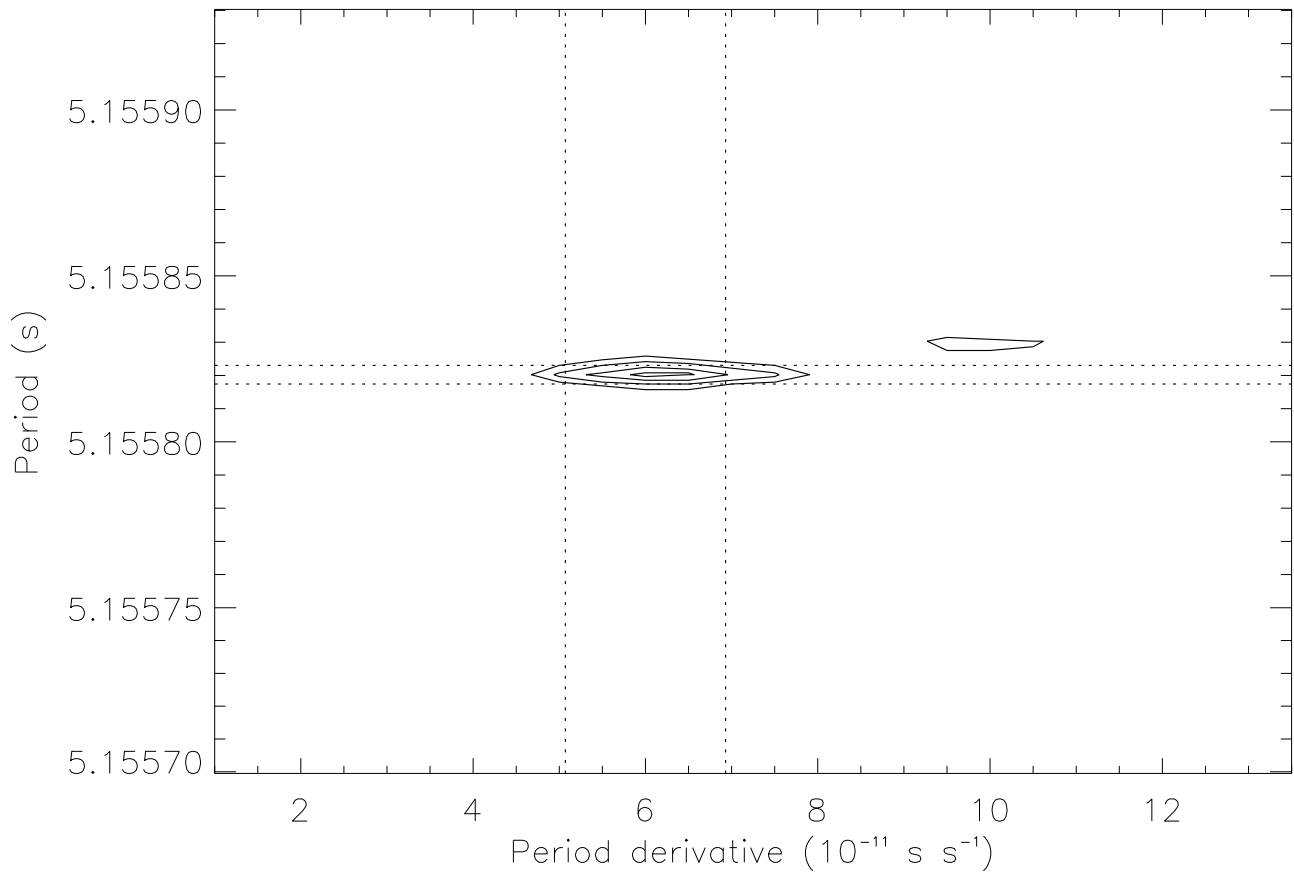
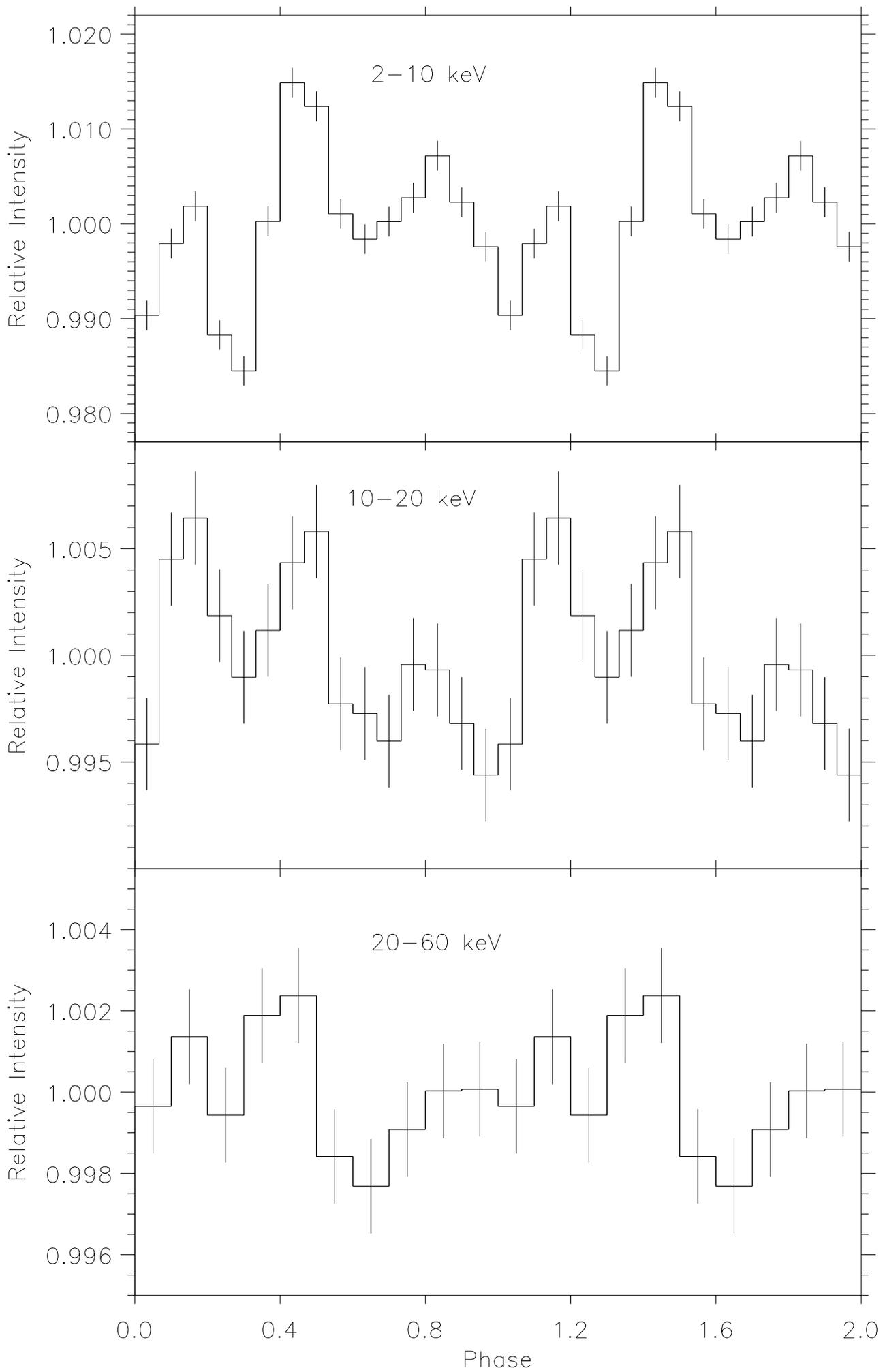
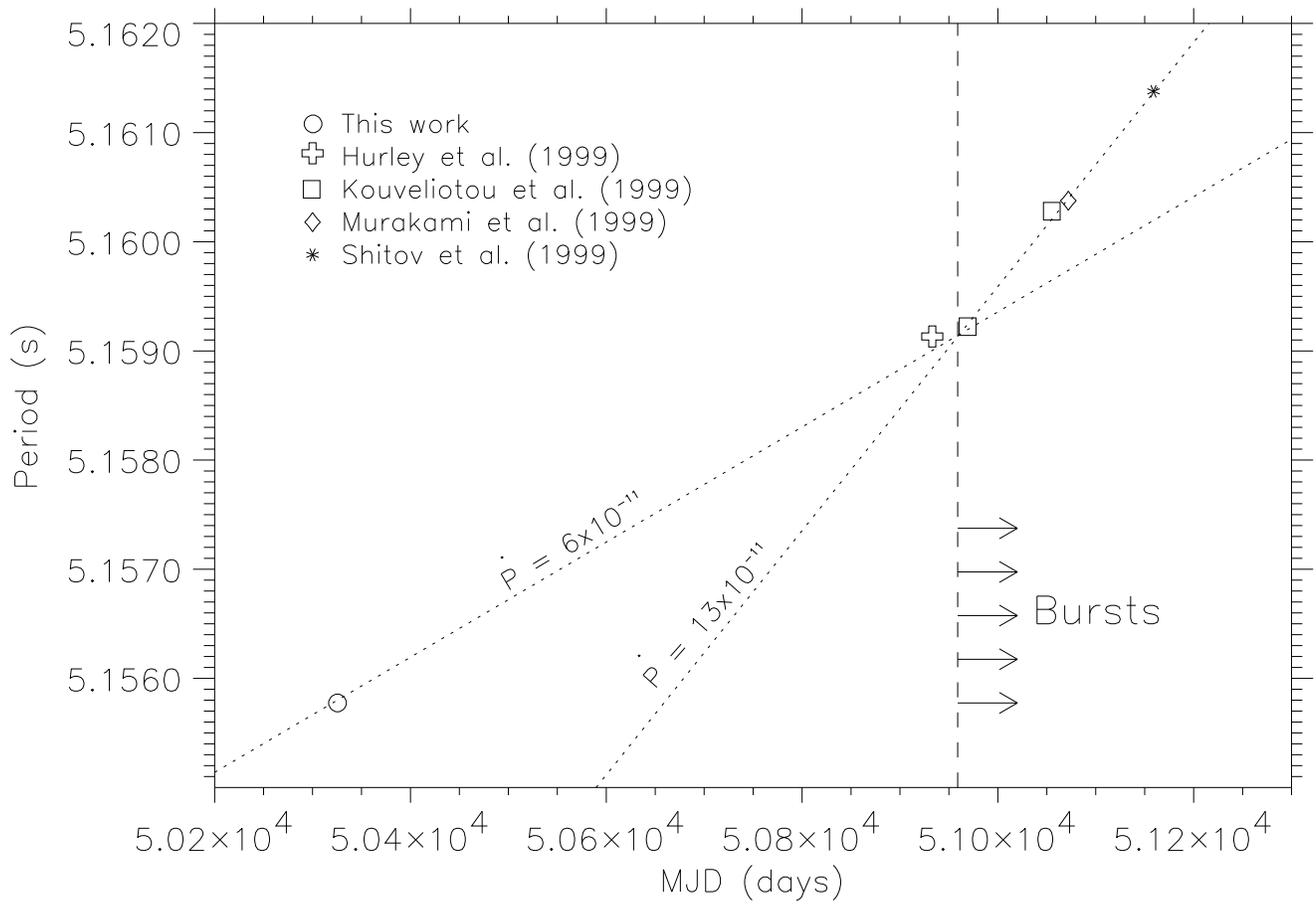
\end{document}